\documentclass[12pt, preprint]{aastex}
\usepackage{natbib}
\usepackage{amsmath}

\begin{document}
\title{Magnetic field intensification by three-dimensional
``explosion'' process}
\author{H. Hotta$^1$, M. Rempel$^2$ and T. Yokoyama$^1$}
\affil{$^1$Department of Earth and Planetary Science, University of Tokyo,
7-3-1 Hongo, Bunkyo-ku, Tokyo 113-0033, Japan\\
$^2$High Altitude Observatory, National Center for Atmospheric Research,
Boulder, CO, USA}
\email{ hotta.h@eps.s.u-tokyo.ac.jp}

\begin{abstract}
We investigate an intensification mechanism for the magnetic field
near the base of the solar convection zone that does not rely on
differential rotation. Such mechanism in addition to differential
rotation has been suggested by studies of flux emergence, which
typically require field strength in excess of those provided by
differential rotation alone. We study here a process in which potential
energy of the superadiabatically stratified convection zone is
converted into magnetic energy. This mechanism, know as "explosion of
magnetic flux tubes", has been previously studied in the thin flux tube
approximation as well as two-dimensional MHD simulations, we expand the
investigation to three-dimensional MHD simulations.
Our main result is that enough intensification can be achieved in a 
three-dimensional magnetic flux sheet as long as the spatial scale of
the imposed perturbation normal to the magnetic field is sufficiently large.
When this spatial scale is small, the flux sheet tends to rise toward
the surface, resulting in a significant decrease of the magnetic
field amplification. 
\end{abstract}

\section{Introduction}\label{introduction}
The solar magnetic field is thought to be generated by dynamo action
\citep{1955ApJ...122..293P}. In most solar dynamo models
differential rotation is the key process that generates strong toroidal
field near the base of the convection zone.
\citep{1995A&A...303L..29C,1999ApJ...518..508D,2010ApJ...709.1009H,2010ApJ...714L.308H}.
The mean energy density of dynamo-generated magnetic field is determined
by the energy density of differential rotation as:
\begin{eqnarray}
 \frac{B^2}{8\pi}\sim\frac{1}{2}\rho(\Delta v)^2,
\end{eqnarray}
where $B$, $\rho$, and $\Delta v$ denote the strength of magnetic field,
density and typical difference of the fluid velocity at the tachocline.
 At the tachocline these values are
$\rho=0.2\ \mathrm{g\ cm^{-3}}$,
$\Delta v=2\pi R_\mathrm{bc}\Delta\Omega=2\pi\times5\times10^{10}\ \mathrm{cm}\times50\
\mathrm{nHz}=15700\ \mathrm{cm\ s^{-1}}$ 
\cite[see the results of helioseismology in][]{2003ARA&A..41..599T}, 
where $R_\mathrm{bc}$ and $\Delta \Omega$ are the radius at the
tachocline and the difference of angular velocity between the top and
the bottom of the tachocline, respectively.
In addition, if we assume a linear velocity profile and formally
integrate that over the width of the shear layer, we have a drop by
another factor of 3 in energy.
Using these values
the dynamo-generated magnetic field is estimated to be approximately
$B\sim1.4\times10^4\ \mathrm{G}$. This estimate is similar to the values found 
in mean-field models with Lorentz force feedback by \citet{2006ApJ...647..662R}.
Note that in \cite{2006ApJ...647..662R}
latitudinal shear mostly generates toroidal magnetic field and that
differential rotation is continuously replenished through the $\Lambda$-effect
(parameterized turbulent angular momentum transport), which explains why 
$10^4$ G can be reached with only small feedback on differential rotation.
\par
On the other hand, theoretical studies of flux emergence and their comparison
with observed sunspot properties (Joy's law, low latitude emergence) imply
substantially stronger field near the base of the convection zone.
\cite{2011ApJ...741...11W} 
suggest that a toroidal field strength of about 50
kG is required at the base of the convection zone in order to reproduce
the statistical properties of tilted sunspot pairs in rising flux
tube simulations \cite[see also review by][]{2009LRSP....6....4F}.
\par
Above results suggest that additional mechanisms for the amplification 
other than rotational shear are required. One candidate of such a 
mechanism is the ``explosion'' of rising magnetic flux tubes.
\citep{1994ApJ...433..867P,1995ApJ...452..894M,2001ApJ...552L.171R}.
It is summarized as follows:
Suppose that there is a superadiabatically stratified atmosphere
like the solar convection zone and a dynamo-generated toroidal magnetic
field. When the gravity has the profile of 
$g(r)\propto r^{-2}$, where $r$
denotes the radius, the distribution of gas pressure $p_\mathrm{e}(r)$ is expressed
as
\begin{eqnarray}
 p_\mathrm{e}(r)=p_\mathrm{base}
\left[
1+\nabla\frac{r_\mathrm{base}}{H_\mathrm{base}}
\left(
\frac{r_\mathrm{base}}{r}-1
\right)
\right]^{1/\nabla},
\end{eqnarray} 
using a constant temperature gradient as
\begin{eqnarray}
 \nabla\equiv \frac{d\log T}{d\log p}.
\end{eqnarray}
$p_\mathrm{base}$ and $H_\mathrm{base}$ denote
the gas pressure and the
pressure scale height at $r=r_\mathrm{base}$, respectively, where
$r_\mathrm{base}$ is the location of the base of the convection zone.
As the toroidal field buoyantly rises as a flux tube, the thermal
property in the tube vary adiabatically while the surrounding gas in the
convection zone is stratified in a superadiabatic manner.
The thermal exchange is mainly by radiative diffusion in
the solar convection zone and the time scale of it is at least several
tens of years for a $1000\ \mathrm{km}$ flux sheet, which is much longer 
than the time scale of flux emergence on the order
of months. Therefore the internal distribution of the gas pressure
is expressed as:
\begin{eqnarray}
 p_\mathrm{i}(r)=p_\mathrm{base}
\left[
1+\nabla_\mathrm{ad}\frac{r_\mathrm{base}}{H_\mathrm{base}}
\left(
\frac{r_\mathrm{base}}{r}-1
\right)
\right]^{1/\nabla_\mathrm{ad}},
\end{eqnarray} 
where $\nabla_\mathrm{ad}$ denotes the adiabatic temperature gradient.
We note that $\nabla>\nabla_\mathrm{ad}$ and 
the pressure of the internal medium of magnetic flux decreases slower
with $r$
than that of the external
medium. Suppose that the pressure is in balance at the base of
convection zone as:
\begin{eqnarray}
p_\mathrm{i}(r_\mathrm{base})+\frac{B^2(r_\mathrm{base})}{8\pi}=p_\mathrm{e}(r_\mathrm{base}).
\end{eqnarray}
As the decrease of the pressure in the tube is slower than that of the surrounding plasma, 
these pressures become the
same at a certain height $r=r_\mathrm{expl}$, i.e.,
$p_\mathrm{i}(r_\mathrm{expl})=p_\mathrm{e}(r_\mathrm{expl})$. This
height is called the explosion height. Even after passing the explosion
height, the flux tube continues to rise up to the surface of the
sun $r=r_\mathrm{t}$. During this process,
mass in the flux tube is swept out due to the imbalance between
the pressure gradient and the gravitational force along the magnetic field 
\cite[see Fig. 2 of][]{2001ApJ...552L.171R}.
We specify the gas pressure distribution $p_\mathrm{i,t}(r)$, after
this sweeping of mass.
Since the gas pressure is balanced at the surface,
i.e. $p_\mathrm{i,t}(r_\mathrm{t})=p_\mathrm{e}(r_\mathrm{t})$,
the pressure distribution is expressed as:
\begin{eqnarray}
p_\mathrm{i,t}(r) = p_\mathrm{e}(r_\mathrm{t})
\left[
1+\nabla_\mathrm{ad}\frac{r_\mathrm{t}}{H_\mathrm{t}}
\left(
\frac{r_\mathrm{t}}{r}-1
\right)
\right]^{1/\nabla_\mathrm{ad}},
\end{eqnarray}
where a subscript ``t'' denotes the value at the surface.
If both ends of the the rising flux tube stay at the initial position
($r=r_\mathrm{base}$), the amplified magnetic
field has the energy density of 
\begin{eqnarray}
 \frac{B^2_\mathrm{amp}}{8\pi}=p_\mathrm{e}(r_\mathrm{base})
  -p_\mathrm{i,t}(r_\mathrm{base}),
  \label{amp}
\end{eqnarray}
Since the pressure is in balance at the surface with the amplified field,
the magnetic field is amplified up to the strength for
which the explosion height approaches the surface. The asymptotic 
field strength is independent of the initial magnetic field and given
by the superadiabaticity of the convection zone. 
\par
\cite{1995ApJ...452..894M} investigate explosion using
one-dimensional thin flux tube approximation.
\cite{2001ApJ...552L.171R} show using two-dimensional MHD calculation
that magnetic field can be amplified up to $10^5\ \mathrm{G}$ in the
solar convection zone.
The time scale of amplification is determined by the
initial magnetic field and is $<0.5\ \mathrm{year}$ when the initial
plasma beta is $\beta<10^7$.\par
In this study, we investigate the possibility of an explosion of a
magnetic flux and its consequential intensification by MHD simulations
in three-dimensions, i.e. in more realistic situations, which have not
yet been investigated in previous studies.

\section{Model}\label{model}
The three-dimensional magnetohydrodynamic equations are solved in
Cartesian coordinates $(x,y,z)$, where $x$ and $y$ denote the horizontal
directions and $z$ denotes the vertical direction.
Equations are expressed as:
\begin{eqnarray}
&& \frac{\partial \rho}{\partial t}=-\nabla\cdot(\rho{\bf v}),\\
&& \frac{\partial }{\partial t}
  \left( \rho{\bf v}\right) = -\nabla\cdot
  \left[\rho{\bf vv}+\left(p+\frac{B^2}{8\pi}\right){\bf I}
  +\frac{\bf BB}{4\pi}\right]-\rho g{\bf e_z},\\
&& \frac{\partial {\bf B}}{\partial t}=\nabla\times({\bf v}\times{\bf B}),\\
&& \frac{\partial e}{\partial t}=\nabla\cdot
 \left[\left( e+p+\frac{B^2}{8\pi}\right){\bf v}
  -({\bf B}\cdot{\bf v})\frac{\bf B}{4\pi}\right]
-\rho gv_z,\\
&&e=\frac{p}{\gamma-1}+\frac{1}{2}\rho v^2+\frac{B^2}{8\pi} \label{total_energy},
\end{eqnarray}
where $\rho$, $p$, ${\bf v}$, ${\bf B}$, $g$ and $e$ denote density,
pressure, fluid velocity, magnetic field, gravitational acceleration,
and total energy per unit volume. The value of a specific heat ration is
set as $\gamma=5/3$. Note that since we use the equation of state
for the perfect gas, the internal energy per unit
volume is expressed as $p/(\gamma-1)$ in eq. (\ref{total_energy})
\par
The divergence free condition, i.e. $\nabla\cdot{\bf B}=0$, is
maintained using the method introduced in
\cite{2002JCoPh.175..645D}.
The simulation domain is 
$(-7.5,-20,-1)<(x/H_\mathrm{b},y/H_\mathrm{b},z/H_\mathrm{b})<(7.5,20,3)$
and the total grid number is $256\times128\times384$.
We use a forth-order centered finite difference scheme for spatial-derivatives 
in combination with a forth-order Runge-Kutta integration in time.
We do not use any explicit viscosity or magnetic diffusivity, but apply
artificial diffusivities to maintain numerical stability
\citep{2009ApJ...691..640R}. We use
periodic boundary conditions for all variables in horizontal
directions. At the top and bottom boundary, stress-free and
non-penetrating boundary conditions are adopted, i.e. 
$\partial v_x/\partial z= \partial v_y/\partial z =0$ and $v_z=0$. The
values of density and pressure at these boundaries are derived from
solving eq. (\ref{entropy_grad})-(\ref{density_trans}). At the top
boundary, the magnetic field is set as 
$B_x=B_y=0$, and $\partial B_z/\partial z=0$. 
At the bottom boundary the magnetic field is set as,
$\partial B_x/\partial z=\partial B_y/\partial z=\partial B_z/\partial
z=0$.
\par
The profile of gravitational acceleration is assumed to be
\begin{eqnarray}
 g(z) = g_\mathrm{b}
\left(\frac{z + r_\mathrm{b}}{r_\mathrm{b}}\right)^2,
\end{eqnarray}
where $r_\mathrm{b}=5H_\mathrm{b}$ is the location of $z=0$
from the center of the gravity source.
We note that the subscript `b' means the value at $z=0$.
Here $H_b$ is used instead of $H_\mathrm{base}$ since the setup of the simulations here
is similar to but not identical with the model in Section \ref{introduction}.
The initial magnetic field is set
as a flux sheet with a perturbation as:
\begin{eqnarray}
&& B_x=B_0f\exp
  \left[-\frac{(z-z_\mathrm{c})^2}{a^2}\right],\\
&& a= a_0 + a_1\exp
 \left[
-\left(\frac{x}{d_x}\right)^2
-\left(\frac{y}{d_y}\right)^2
\right],\\
&& z_\mathrm{c}= z_0 + z_1\exp
 \left[
-\left(\frac{x}{d_x}\right)^2
-\left(\frac{y}{d_y}\right)^2
\right],\\
&& f= \frac{a_0}{a},\\
 && B_0 = \sqrt{\frac{8\pi p_\mathrm{b}}{\beta}}, \\
\end{eqnarray}
where $a$, and $z_\mathrm{c}$ denote the thickness and the
center of a flux sheet, respectively. $d_x$ and $d_y$ are the
range of perturbation in $x$ and $y$ direction, respectively.  We
specify $a_0=0.2H_\mathrm{b}$, $a_1=0.25H_\mathrm{b}$,
$z_0=0.1H_\mathrm{b}$, $z_1=0.25H_\mathrm{b}$, $\beta=300$ and
$d_x=2H_\mathrm{b}$. $d_y$ is a free parameter from $2H_\mathrm{b}$
to $8H_\mathrm{b}$. 
As shown above, we are considering a toroidal flux sheet as a source of
the magnetic flux prior to the intensification because the magnetic
field stretched by the $\Omega$-effect at the bottom of the convection zone
is suggested to have a sheet structure according to
\cite{2006ApJ...647..662R}.
Then $B_z$ is calculated to satisfy the divergence free condition,
i.e. $\partial B_x/\partial x+\partial B_z/\partial z=0$ with $B_z=0$ at
the all boundaries and $B_y=0$ in all the domain.\par
In order to mimic the situation of the solar convection zone, we adopt
a similar approach as \cite{2001ApJ...552L.171R}. The upper layer ($z>0$) is 
adiabatically stratified (convection zone: $\delta=0$) and the
lower layer ($z<0$) is
subadiabatically stratified (radiative zone: $\delta=-0.2$), where
$\delta$ is the superadiabaticity. The magnetic layer is placed at the
interface between both regions. Since our convection zone is adiabatically
stratified we add additional entropy within the flux sheet to allow for a 
buoyant rise and subsequent explosion. Using this setting, the pressure scale
height in the flux sheet is
longer than that of the external medium and the explosion occurs around 
$z_\mathrm{expl}\sim H_\mathrm{b}c_\mathrm{p}/\beta \Delta s$, where
$\Delta s$ is the added entropy. 
Using an adiabatic stratification in the upper layer allows us to study
the explosion process in a relatively simple situation without thermal
convection; this leads, however, also to limitations since some key processes
are not catured in our setup. Beside not capturing the influence from
convection, our magnetic flux sheet is not stored in stable layer having
neutral magnetic buoyancy, i.e. overshoot region at the base of the solar 
convection zone. The flux sheet is however stabilized against buoyancy to some
degree by lying on top of our stably stratified lower layer. 
We will investigate the explosion process in a superadiabatic convection zone 
with fully developed thermal convection in a future study.

The initial conditions for pressure
and density are derived from solving the equations:
\begin{eqnarray}
&&\frac{\partial s}{\partial z}=-\frac{c_\mathrm{p}\delta}{H_\mathrm{p}}
 -\frac{2\Delta s(z-z_\mathrm{c})}{a^2}
\exp\left[-\frac{(z-z_\mathrm{c})^2}{a^2}\right]\label{entropy_grad},\\
&& \frac{\partial p_\mathrm{tot}}{\partial z} = -\rho g,\\
&& p_\mathrm{tot} = p + \frac{B^2}{8\pi},\\
&& \rho = \left[\frac{p}{\exp(s/c_\mathrm{v})}\right]^{1/\gamma}\label{density_trans}.
\end{eqnarray}
We specify $\Delta s= 0.03c_\mathrm{v}$. Using this parameter the
explosion height is around $z_\mathrm{expl}=0.18H_\mathrm{b}$.
We do not yet aim at a realistic simulation with solar parameters but
rather consider a model that allows us to study the explosion and
the following intensification in (artificial) isolation.
If there is a
superadibatically stratified convection zone, turbulent thermal
convection is generated and the situation becomes more complicated.\par
\section{Result}\label{result}
\begin{figure}[htbp]
 \centering
 \includegraphics[width=15cm]{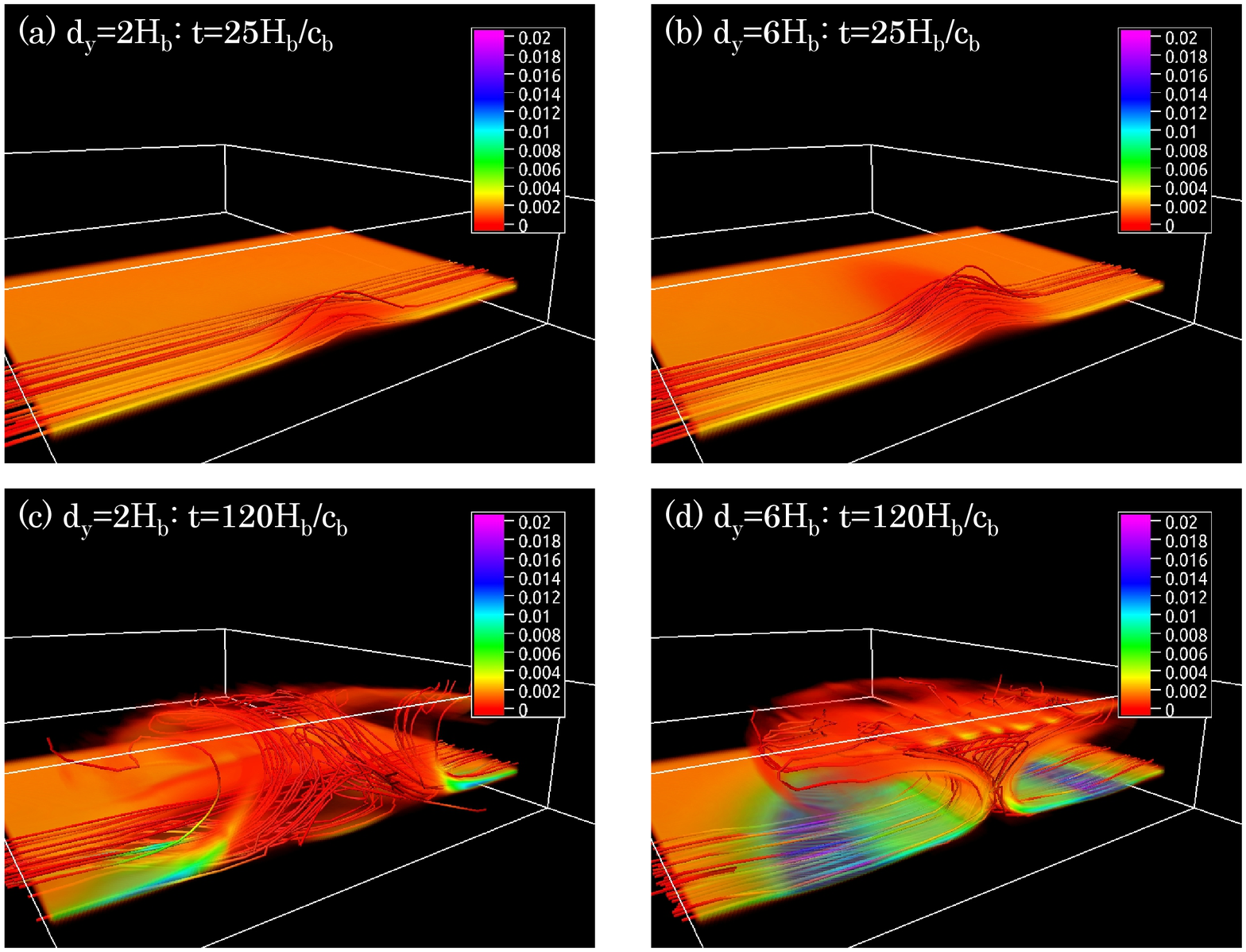}
 \caption{The volume-rendered magnetic energy and the
 magnetic field lines. The panels a and c (b and d) show the results
 with $d_y=2H_\mathrm{b}$ ($d_y=6\mathrm{b}$).
 For convenience of presentation, we show only the solution in the 
center region cut at $y=0$. The color shows the magnetic energy
 normalized by the gas pressure at $z=0$ ($p_\mathrm{b}$).
 This visualization is created using the software of VAPOR
 \citep{Clyne:VDA:2005,Clyne:NJP:2007}.\label{3d}}
\end{figure}

In this study, we take the $y$-component of the perturbation on the
initial magnetic field ($d_y$) as a free parameter.
Fig. \ref{3d} shows the strength and the field lines of the magnetic
field for simulations with $d_y=2H_\mathrm{b}$ and $d_y=6H_\mathrm{b}$.
We present here only the solutions in the center region cut at
$y=0$ to enhance the clarity of the presentation. The magnetic layer fills 
in the horizontal directions the whole domain outlined by the white lines.
In our three-dimensional situation a value of $d_y=6H_\mathrm{b}$
(Fig. \ref{3d}d) leads to a substantial intensification of magnetic field, 
which is similar to the two-dimensional calculation reported by
\cite{2001ApJ...552L.171R}. At the foot-points the magnetic field lines 
remain almost parallel to the $x$-axis for $d_y=6H_\mathrm{b}$.
On the other hand, the amplification of magnetic field is significantly
less with $d_y=2H_\mathrm{b}$.
In the result with $d_y=2H_\mathrm{b}$, the foot-point magnetic field
rises upward and the magnetic field lines at the back ($y<0$) are bent
to the center ($y\sim0 $) of the calculation domain.
At this stage, it is suggested that the
intensification of magnetic field by explosion process is more
likely to occur with a longer perturbation, which is perpendicular to
the main magnetic field.\par
\begin{figure}[htbp]
 \centering
 \includegraphics[width=15cm]{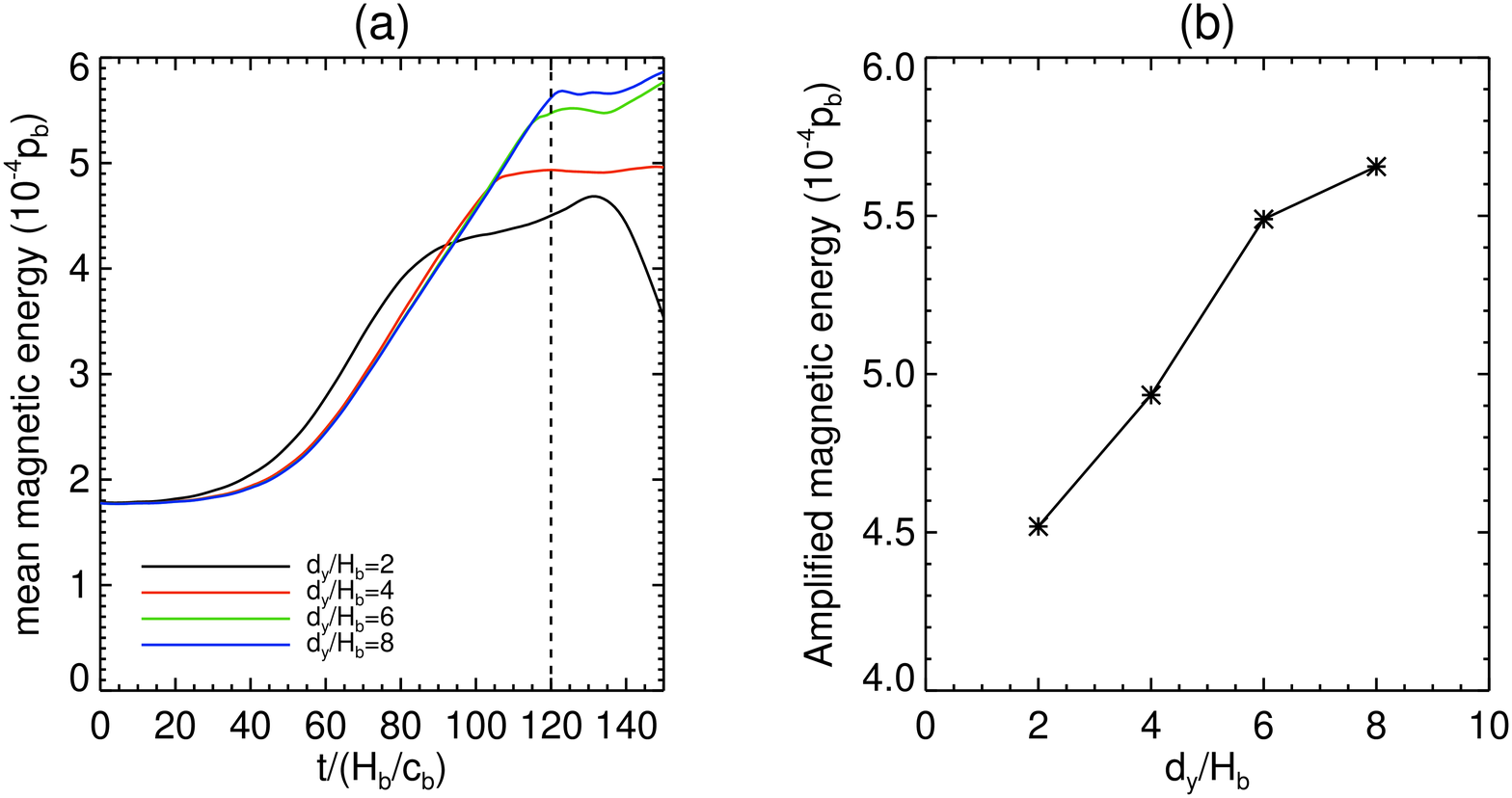}
 \includegraphics[width=15cm]{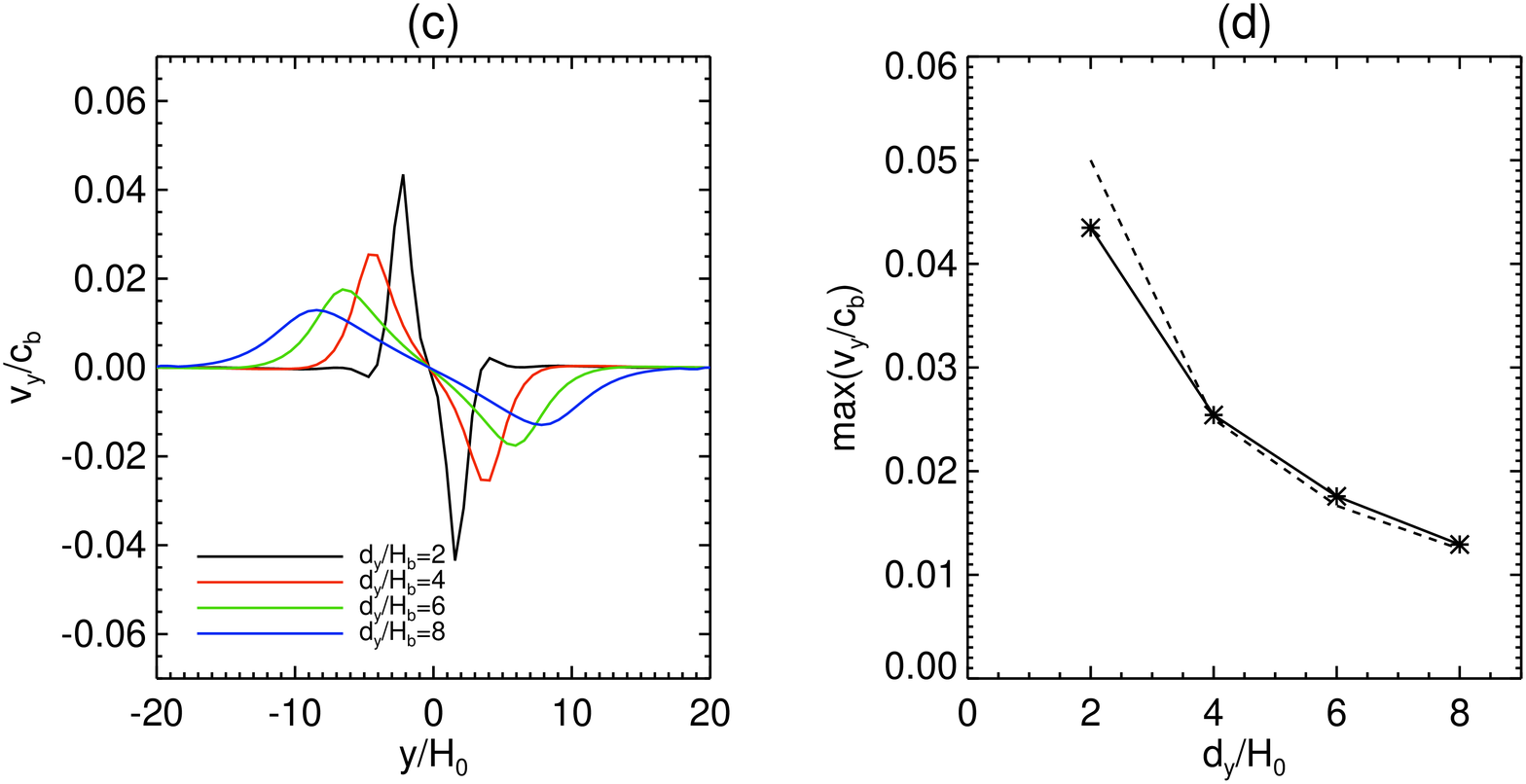}
 \caption{(a) Temporal evolution of the mean magnetic energy
 $|B^2/8\pi|$ on the $y=0$
 plane. (b) The dependence of the mean magnetic energy at
 $t=120H_\mathrm{b}/c_\mathrm{b}$ on $d_y$. The energy of magnetic field
 is  normalized by $10^{-4}p_\mathrm{b}$.
(c) The profile of $y$-component fluid velocity averaged over
 $-0.1H_\mathrm{b} < x <0.1H_\mathrm{b}$ and $0<z<2H_\mathrm{b}$ at
 $y=0$. (d) The dependence of the maximum value of the averaged
 $y$-component velocity on $d_y$. The dashed line denotes the dependence
 of $v_y=0.1/d_y$.
\label{meanb}}
\end{figure}

Fig. \ref{meanb}a shows the temporal evolution of magnetic energy
averaged over the $y=0$ plane. Around $t=120H_\mathrm{b}/c_\mathrm{b}$,
where $c_\mathrm{b}$ is the speed of sound at $z=0$, the initial 
intensification of magnetic field with explosion process is completed 
in all cases, i.e. the top of the rising flux is reaching the upper boundary. 
Fig. \ref{meanb}b
shows the dependence on $d_y$ of the mean magnetic energy at
$t=120H_\mathrm{b}/c_\mathrm{b}$. It is qualitatively clear that
a larger scale $d_y$ causes a more effective intensification. 
\par
\begin{figure}[htbp]
 \centering
 \includegraphics[width=15cm]{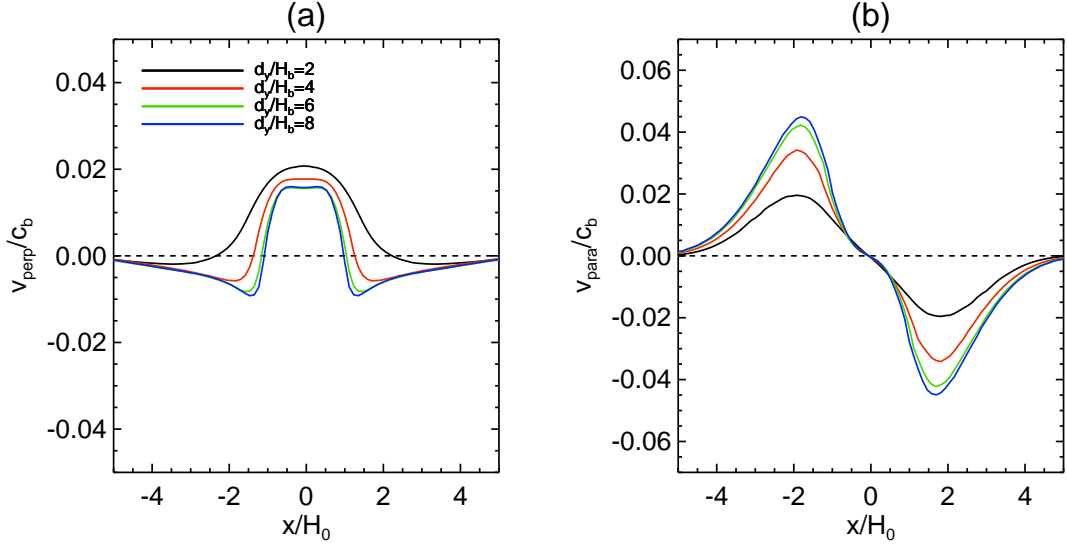}
 \caption{(a) Averaged $v_\mathrm{perp}$ along magnetic field
 lines, where $v_\mathrm{perp}$ is the fluid speed perpendicular to the
 magnetic field.
 We choose the magnetic field lines which
 start $x=y=0$ and $z=0$, $0.25H_\mathrm{b}$.
 (b) Averaged velocity parallel to magnetic field $v_\mathrm{para}$ is
 shown. The same magnetic field lines for the panel (a) are choosen.
\label{tilt}}
\end{figure}

In the following discussion, we compare the results with different $d_y$ at the
time when the apex of the rising magnetic flux sheet reaches $z=2H_\mathrm{b}$ 
in each case. Fig. \ref{meanb}c shows the
profile of the $y$-component of the fluid velocity averaged over 
$-0.1H_\mathrm{b}<x<0.1H_\mathrm{b}$ and $0<z<2H_\mathrm{b}$ at $y=0$.
Fig \ref{meanb}d shows the dependence of the maximum value of
the averaged $y$-component fluid velocity. The flow parallel to the
$y$-axis is generated by a pressure gradient $\partial p/\partial y$: As the
magnetic flux rises, the gas pressure in the central region decreases and
it causes a converging flow in the $y$-direction. The amplitude of the fluid 
velocity is estimated as $v_y=\tau\Delta p/(\rho d_y)$, where $\Delta p$ 
denotes the perturbations of the gas pressure with respect to the ambient 
medium and $\tau$ the time intervall over which the acceleration takes place.
For our estimate we use values of
$\rho\sim\rho_\mathrm{b}$, and $\tau\sim100H_\mathrm{b}/c_\mathrm{b}$,
The difference of the pressure is estimated to be $\Delta
p\sim0.001p_\mathrm{b}$ which is the same order of
$p_\mathrm{b}/\beta$, 
where $\rho_\mathrm{b}$ and $p_\mathrm{b}$ are the
density and pressure at $z=0$.
Thus we obtain $v_y=0.1H_\mathrm{b}c_\mathrm{b}/d_y$. 
The dashed line in Fig. \ref{meanb}d corresponds to this equation.
Overall the result with $d_y=4H_\mathrm{b}$, $6H_\mathrm{b}$, and
$8H_\mathrm{b}$ are well reproduced. The result with $d_y=2H_\mathrm{b}$ is 
slightly smaller
than the value estimated from $v_y=0.1/d_y$. The reason is that the
center region is already filled with mass, resulting in a smaller
$\Delta p$. In addition, a small value of $\tau$ with
$d_y=2H_\mathrm{b}$ causes the deviation.
Fig. \ref{tilt}a shows the value of
$v_\mathrm{perp}$ along the field lines, where $v_\mathrm{perp}$ is
the fluid velocity perpendicular to the magnetic field in the $x$-$z$ plane.
A positive (negative) value of $v_\mathrm{perp}$ indicates
a flow which moves the magnetic flux upward
(downward). 
Using small
(large) $d_y$, the foot-point magnetic flux move upward (downward).
Fig \ref{tilt}b shows the velocity parallel to the magnetic field along
magnetic field lines. The larger velocity is generated with larger
$d_y$. These correspond to the amplification of the magnetic field.
\par
\section{Discussion}
\begin{figure}[htbp]
 \centering
 \includegraphics[width=15cm]{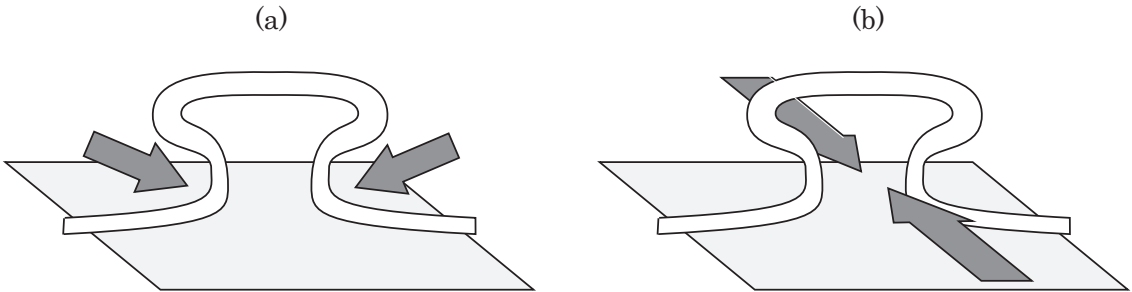}
 \caption{Schematic picture of the explosion process in
 three-dimensional situation with (a) large and (b) small $d_y$.
 \label{schema}}
\end{figure}

\begin{figure}[htbp]
 \centering
 \includegraphics[width=7.5cm]{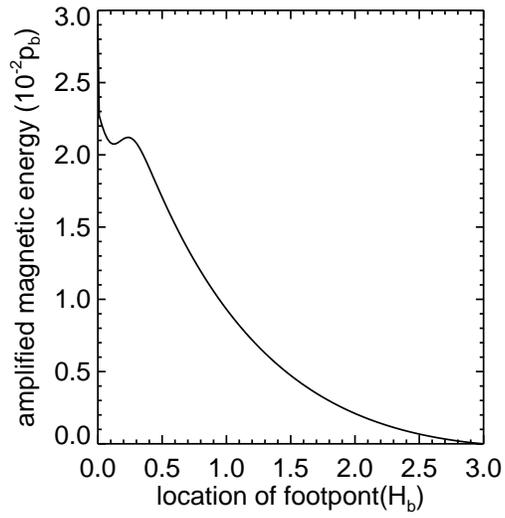}
 \caption{
The dependence of the amplified magnetic energy by the
 explosion process on the location of foot-point.\label{anal}}
\end{figure}

Our results show that the efficiency of the magnetic intensification is
dependent on the spatial scale of the perturbations ($d_y$). The reason is
understood as follows:
In a two-dimensional situation \citep{2001ApJ...552L.171R} or using
large value of $d_y(>6H_\mathrm{b})$, the pressure gradient
$\partial p/\partial x$ leads to strong converging flows toward the center
of the rising flux sheet within the $y=0$-plane. The flow component parallel to
the field leads to amplification, the flow component perpendicular to the
field prevents the buoyant motion around the legs of the rising flux and
consequently plays a role for pinning down the foot-points
(Fig. \ref{schema}a).
On the other hand, using
small $d_y(<2H_\mathrm{b})$,
a converging motion in the $y$-direction
is promoted by the large pressure gradient
$\partial p/\partial y$ and the center of the explosion region is filled
with the mass. Thus the pressure around the foot-point becomes large and
the flow $v_x$ is suppressed.
Consequently magnetic flux moves upward and is less amplified 
(Fig. \ref{schema}b).
If the foot-point magnetic field is fixed at the initial position,
the amplification of magnetic energy is estimated with eq. (\ref{amp}).
In a three-dimensional situation, however there is a possibility that the
foot-point magnetic flux moves upward, thus the amplified magnetic energy
is estimated using eq. (\ref{amp}) at $r=r_\mathrm{f}$ instead of
$r=r_\mathrm{base}$,
where $r_\mathrm{f}$ denotes the location of the foot-point magnetic
flux. Fig. \ref{anal} shows the dependence of the amplified magnetic
energy by the explosion process on the location of foot-point with the
parameters of our calculation.
If the foot-point magnetic
field is fixed in the initial position, the magnetic energy of
$\sim2\times10^{-2}p_\mathrm{b}$ can be generated by the explosion
process. The result with large $d_y(=6H_\mathrm{b})$ well reproduces this
estimation (see Fig. \ref{3d}d).
Note that Fig. \ref{meanb}a shows the mean magnetic energy in the 
$y=0$ plane, leading to a lower values of $\sim 6\times10^{-4}p_\mathrm{b}$ 
at maximum.
Fig. \ref{anal} shows that the efficiency
of the intensification significantly decrease as the foot-point magnetic
field rises.
It is confirmed that even using larger $\beta(=600)$, we can obtain
similar results, i.e. strong
amplification is achieved with larger $d_y(>6H_\mathrm{b})$.
\section{Summary}
We investigate an intensification mechanism of magnetic field near the 
base of solar convection zone following the ``explosion'' process using
three-dimensional MHD simulation. Such a mechanism in addition to differential
rotation has been suggested by studies of flux emergence, which
typically require field strength in excess of those provided by
differential rotation alone.
Our main result is that the enough
intensification can be achieved even in a three-dimensional situation 
as long as the spatial scale of the
imposed perturbation in the direction normal to the magnetic flux sheet is
sufficiently large.
When this spatial scale is small,
the efficiency of intensification of magnetic
field significantly decreases. Here a strong converging flow in the direction
perpendicular to the field ($y$-direction) is driven, which leads to weaker 
flows in the $x$-direction. The latter are crucial for the field amplifiction
as well as stabilization of the foot-points in our setup.
\par
In this study, we used a similar setup as \cite{2001ApJ...552L.171R}, which
allows to study the explosion process in an adiabatically stratified 
atmosphere, avoiding the complex situation caused by presence of turbulent 
thermal convection. Using a more realistic setup
with a superadiabatic convection zone and storage of magnetic flux in a 
subadiabatic region beneath could influence our results in the following 
way:

If the initial magnetic flux is stored in a subadiabatic stratification the 
foot-point magnetic field is more stable and the criterion for the length
scale of the perturbation might be relaxed, i.e. also shorter perturbations
could lead to stronger amplification. However, even with more realistic 
storage in a subadiabatic region, there
is no force that could stop the horizontal motion of the foot-points
(buoyancy only works in vertical direction). As a consequence also here a
fundamental difference between tube-like and sheet-like behavior will
persist.

The presence of turbulent thermal convection could significantly influence
our results since during the explosion process the strength of magnetic 
field near the top of the rising flux drops significantly. Thus, it is 
possible that the connection between top and foot-point
magnetic field is disturbed by turbulence, resulting in a reduction 
of field intensification. 

Therefore more realistic simulations including a stably stratified overshoot 
region/radiation zone and a superadiabatic convection zone are required in
order to investigate the detailed conditions under which the explosion 
process leads to substantial intensification of magnetic field.

\acknowledgements
 The authors are grateful to Y. Fan for her helpful comments.
 Numerical computations were, in part, carried out on a Cray XT4 at the Center
 for Computational Astrophysics, CfCA, of the National Astronomical
 Observatory of Japan.
 This work was supported by Grant-in-Aid for JSPS Fellows.
 This work was supported by the JSPS Institutional Program for
 Young Researcher Overseas Visits and the Research
 Fellowship from the JSPS for Young Scientists.
 The National Center for Atmospheric Research is sponsored by the
 National Science Foundation.



\end{document}